\documentstyle[prl,aps,epsfig]{revtex}
\tighten
\newcommand{\ovl}{\overline}
\newcommand{\nn}{\nonumber}

\newcommand{\al}{\alpha}

\newcommand{\de}{\delta}

\newcommand{\La}{\Lambda}

\newcommand{\Om}{\Omega}

\newcommand{\th}{\theta}

\newcommand{\ra}{\rightarrow}

\newcommand{\be}{\begin{equation}}
\newcommand{\ee}{\end{equation}}

\newcommand{\bea}{\begin{eqnarray}}
\newcommand{\eea}{\end{eqnarray}}
\newcommand{\bean}{\begin{eqnarray*}}
\newcommand{\eean}{\end{eqnarray*}}
\newcommand{\dd}{\partial}

\begin{document}
\draft
\preprint{\ 
\begin{tabular}{rr}
UGVA-DPT 1999/11-zzzz &  \\ 
astro-ph/0102144 & 
\end{tabular}
} 
\title{Dynamical Instabilities of the Randall--Sundrum Model} 
\author{Timon Boehm$^1$, Ruth Durrer$^1$ and Carsten van de Bruck$^2$}
\address{$^1$D\'epartement de Physique Th\'eorique, Universit\'e de
Gen\`eve, 24 quai E. Ansermet, CH--1211 Geneva 4 (Switzerland).}
\address{$^2$Department of Applied Mathematics and Theoretical Physics, 
University of Cambridge, Wilberforce Road, \\ Cambridge CB3 0WA, UK}

\date{\today}
\maketitle
\begin{abstract}

We derive dynamical equations to describe a single 3-brane containing
fluid matter and a scalar field coupling to the 
dilaton and the gravitational field in a five dimensional bulk.
First, we  show that a scalar field or an arbitrary fluid on the brane
cannot evolve to cancel the cosmological constant in the bulk. 
 Then we show that the Randall--Sundrum model is unstable
under small deviations from the fine--tuning between the brane tension and 
the bulk cosmological constant and even under homogeneous gravitational 
perturbations. Implications for brane world cosmologies are  discussed. 

\end{abstract}

Preprint: DAMTP-2001-16

\section{Introduction}
Until now, string theories are the most promising fundamental quantum
theories at hand which include gravity. Open strings carry gauge
charges and end on so-called D$p$--branes, $(p+1)$--dimensional
hypersurfaces of the full spacetime. Correspondingly, gauge fields may 
propagate only on the $(p+1)$--dimensional brane, and only modes associated
with closed strings, such as the graviton, the dilaton and the axion, 
live in the full spacetime~\cite{Witten}. Superstring theories and 
especially M theory suggest that the observable universe is a 
$(3+1)$--dimensional hypersurface, a $3$--brane, in a $10$ or $11$
dimensional spacetime. This fundamental spacetime could be 
a product of a four--dimensional Lorentz manifold with an
$n$--dimensional compact space of volume $V_n$ ($n$ being the number of 
extra dimensions). Then, the relation between the $(4+n)$--dimensional
fundamental Planck mass $M_{\mbox{\tiny{f}}}$ and the effective four 
dimensional Planck mass 
$M_{\mbox{\tiny{eff}}}\equiv \sqrt{1/(8\pi G_N)} \simeq  2.4 \times
10^{18}$ GeV is 
\be
M_{\mbox{\tiny{eff}}}^2 = M_{\mbox{\tiny{f}}}^{n+2} V_n.
\ee
If some of the extra dimensions are much larger than the fundamental
Planck scale, $M_{\mbox{\tiny{f}}}$ is much smaller than
$M_{\mbox{\tiny{eff}}}$ and may even be close to the electroweak
scale, thereby relieving the long-standing hierarchy problem~\cite{Hamed}. For example, if one allows for two `large' extra
dimensions of the order of $1\,$mm, one obtains a fundamental Planck
mass of $1\,$TeV. However,
a new hierarchy between the electroweak scale and the mass scale
associated with the compactification volume $\frac{1}{{V_n}^{1/n}}$
is introduced. 

Clearly, this idea is very interesting from the point of view of
bringing  together fundamental theoretical high energy
physics and experiments, which have been diverging more and more since the
advent of string theory. While the four dimensionality of gauge
interactions has been tested down to scales of about $1/200
\,$GeV$^{-1} \simeq 10^{-15}\,$mm, Newton's law is experimentally
confirmed only above $1 \,$mm. Therefore, `large' extra dimensions
are not excluded  and should  be tested in the near future by refined
microgravity experiments~\cite{expG}. The 
fundamental string scale might in principle be accessible to the CERN Large
Hadron Collider  (LHC)~\cite{exp}.

In the past, it was commonly assumed that the fundamental spacetime is
factorizable, and that the extra--dimensional space is compact. Recently, 
Randall and Sundrum~\cite{RSII} proposed a five--dimensional model, in 
which the metric on the $3$--brane is multiplied by 
an exponentially decreasing `warp' factor such that transverse lengths
become small already at short distances along the fifth
dimension. This idea allows for a noncompact extra dimension
without getting into conflict with observational facts. 
In this scenario the brane is embedded in an anti--de Sitter space, 
and a fine--tuning relation 
\begin{equation}\label{fine--tuning}
\La = -\frac{{\kappa_5}^2}{6}V^2, 
\end{equation}
between the brane tension $V$ and the negative cosmological constant in 
the bulk $\La$ has to be satisfied. Here, $\kappa_5$ is related to the
five--dimensional Newton constant by ${\kappa_5}^2 = 8\pi G_5 = {M_5}^{-3}$. 
Randall and Sundrum also proposed a model with two branes of opposite
tension which  provides an elegant way to relieve both hierarchy
problems
mentioned above~\cite{RS}. However, also  this model requires the
fine--tuning (\ref{fine--tuning}).  
Here, we will only consider the case of a single brane.  

The main unattractive feature of the Randall--Sundrum (RS) model is the
fine-tuning condition (\ref{fine--tuning}). Both from the 
particle physics and the cosmological point of view this relation between two 
a priori independent quantities appears unlikely. 
One would like to put it on a physical basis, such as a fundamental principle, 
or explain it due to some dynamical process.  

The purpose of this paper is to point out the cosmological 
problems associated with the fine--tuning condition
(\ref{fine--tuning}). The outline of the paper is as follows: In
Sec.~\ref{sec:equationsofmotion}, we derive dynamical equations 
describing the gravitational field and the dilaton in the bulk
coupling to fluid matter and a scalar field on the brane. These 
equations allow for a dynamical generalization of the RS model, 
which is a special static solution of our equations with vanishing 
dilaton. Our equations also provide a starting point for further
studies of various issues in cosmology, for example inflation.  In  
Sec.~\ref{sec:dynamicalbrane} we discuss a cosmological 
version of the RS model and show that the fine--tuning condition 
(\ref{fine--tuning}) cannot be stabilized by an arbitrary scalar 
field or fluid on the brane. In Sec. IV we discuss linear 
perturbations of the static RS model  and
derive gauge invariant perturbation equations from our general 
setup. We prove that the full RS spacetime is unstable against 
homogeneous processes on the brane such as
cosmological phase transitions: The solutions run  
quadratically fast away from the static RS spacetime. This instability 
reminds that  of the static homogeneous and isotropic Einstein 
universe~\cite{Einstein}. In linear perturbation theory we also find a mode
which represents an instability linear in time. 
In the last section we present our results and the conclusions. 


\section{Equations of motion}
\label{sec:equationsofmotion}
In this section we derive the equations of motion. 
For generality and for future work we have included the dilaton, although
it does not play a role in the present discussion of RS stability.
Works on dilaton gravity and the brane world have also been done by the 
authors of Refs.~\cite{mennim} and \cite{wands}.

\subsection{General case}
We consider a five--dimensional spacetime with  metric $g_{MN}$ parametrized by coordinates 
$(x^M) = (x^{\mu},y)$, where $M = 0, 1, 2, 3, 5$ and  $\mu = 0, 1, 2,
3$, with a $3$--brane fixed at $y=0$. We use units in which 
$2 {\kappa_5}^2 = 1$. In the string frame the action is  
\bea
S_{\tiny{\mbox{string}}} & = & \int d^{5}x \sqrt{-g} e^{-2 \phi} \left(R 
+ 4 (\nabla_M \phi)(\nabla_N \phi) g^{MN}  - \La(\phi) \right) \\ \nn & 
- & \int d^{4}x \sqrt{-\ovl{g}} e^{-2 \phi} 
\left(\frac{1}{2} (\ovl{\nabla}_\mu \varphi)
 (\ovl{\nabla}_\nu \varphi) \ovl{g}^{\mu\nu} + V(\varphi) 
+ \mathcal{L}_{\tiny{\mbox{fluid}}}\right), 
\eea
which describes the dilaton, $\phi$, coupling to gravity as 
well as to a scalar field $\varphi$ with a potential $V(\varphi)$ and to a fluid. The graviton,  dilaton and  `bulk potential' $\La(\phi)$ live 
in the five dimensional (bulk) spacetime, whereas the fluid 
and the scalar field are confined to 
the brane. The induced four--dimensional metric is\footnote{Where confusion could arise, we overline four dimensional quantities.} 
\begin{equation}
\ovl{g}_{\mu\nu}=\de^M_\mu \de^N_\nu g_{MN}(y=0).
\end{equation}
The action in the Einstein--frame is obtained by the  conformal transformation
\be
g_{MN} \rightarrow e^{-\frac{4\phi}{D-2}} g_{MN}
\ee
with $D=5$. We find
\bea
S_{\tiny{\mbox{Einstein}}} & = & \int d^{5}x \sqrt{-g} \left(R 
- \frac{4}{3} (\nabla_M \phi) (\nabla_N \phi) g^{MN}  
- e^{(4/3) \phi} \La(\phi)\right) \\ \nonumber 
& - &  \int d^{4}x \sqrt{-\ovl{g}} \left(\frac{1}{2} e^{-(2/3) \phi} 
(\ovl{\nabla}_\mu \varphi) (\ovl{\nabla}_\nu \varphi) \ovl{g}^{\mu\nu} 
+ e^{(2/3) \phi} \left(V(\varphi) 
+ \mathcal{L}_{\tiny{\mbox{fluid}}} \right) \right),
\eea
where $g$ now denotes the metric tensor in the Einstein--frame, and
$R, \nabla$ and $\ovl{\nabla}$ are constructed from $g$. The equations of motion are obtained by varying this action with respect to the dilaton, 
the brane scalar field and the metric  
\bea
& & \frac{8}{3}\nabla^2 \phi - \frac{4}{3} e^{(4/3) \phi}\La(\phi) 
    - e^{(4/3) \phi} \frac{\dd \La(\phi)}{\dd \phi} + 
    \frac{\sqrt{-\ovl{g}}}{\sqrt{-g}} \de(y) \left(\frac{1}{3} e^{-(2/3) 
    \phi} (\ovl{\nabla}\varphi)^2 - \frac{2}{3}e^{(2/3) \phi}(V(\varphi) +  
    \mathcal{L}_{\tiny{\mbox{fluid}}})\right)=0, \\
& & e^{-(2/3) \phi} (\ovl{\nabla}^2 \varphi) - e^{(2/3) \phi} 
    \frac{\dd V(\varphi)}{\dd \varphi}=0,\\ 
& & G_{MN} = \frac{4}{3}\left((\nabla_M \phi) (\nabla_N \phi) -\frac{1}{2} 
              g_{MN} (\nabla \phi)^2 \right) - \frac{1}{2} g_{MN} e^{(4/3) 
              \phi} \La(\phi) \\ \nn  & - & \frac{\sqrt{-\ovl{g}}}{\sqrt{-g}} 
              \de(y) \de_M^\mu \de_N^\nu \left(-e^{-(2/3) \phi} \frac{1}{2} 
              \left((\ovl{\nabla}_{\mu} \varphi)(\ovl{\nabla}_{\nu} \varphi) 
              -\frac{1}{2} 
              \ovl{g}_{\mu \nu} (\ovl{\nabla} \varphi)^2 \right) + \frac{1}{2} 
              \ovl{g}_{\mu \nu} e^{(2/3) \phi} V(\varphi) 
              - \frac{1}{2} e^{(2/3) \phi} 
              (T_{\tiny{\mbox{fluid}}})_{\mu \nu} \right),
\eea 
where $G_{MN}$ is the five--dimensional Einstein tensor of the metric $g_{MN}$.
\\ \\
As we are interested in cosmological solutions, we require the $3$--brane to be homogeneous and isotropic and make the ansatz
\be
\label{metricansatz}
ds^2 = -e^{2 N(t,y)}dt^2 + e^{2 R(t,y)}d\vec{x}^2 + e^{2 B(t,y)}dy^2~,
\ee
where the ordinary spatial dimensions are assumed to be flat. Note that
this metric is not factorizable as the scale factor on the brane
$e^{R(t,y)}$ and  the lapse function  $e^{N(t,y)}$  depend on time 
as well as on the fifth dimension. The factor 
$e^{B(t,y)}$ is a modulus field. 
The energy--momentum tensor of a homogeneous and isotropic fluid, representing
matter in the universe, is 
\be
\left({{T_{\tiny{\mbox{fluid}}}}^{\mu}}_{\nu}(t)\right)
=\mbox{diag}(-\rho(t), p(t), p(t), p(t)), 
\ee
and for the dilaton and the brane scalar field we shall assume
$\phi = \phi(t,y), \varphi = \varphi(t)$. 
Finally, the Lagrangian density of the fluid, 
$\mathcal{L}_{\tiny{\mbox{fluid}}}$, 
is given by its free energy density $F$ (see Ref. \cite{tseytlin}).
\\ \\
With these assumptions the equations of motion take the following form.
(An overdot and a prime refer to the derivatives with respect to $t$ and $y$, 
and  quantities on the brane  carry a subscript zero, for example $N_0 \equiv N(t,y=0)$.)
\\ \\
\bea
\phi:\;\;\; & & \frac{8}{3}e^{-2N}(\ddot{\phi}-\dot{\phi}\dot{N}+
                      3\dot{\phi}\dot{R}+\dot{\phi}\dot{B})-\frac{8}{3}
                      e^{-2B}(\phi''+\phi'N'+3\phi'R'-\phi'B')+
                      \frac{4}{3}e^{(4/3) \phi}\La(\phi)  \nn \\ 
                  &+& e^{(4/3) \phi} \frac{\dd \La(\phi)}{\dd \phi}+
                      \de(y)e^{-B}\left(\frac{1}{3}e^{-(2/3) \phi}
                      e^{-2N}\dot{\varphi}^2 + 
                      \frac{2}{3}e^{(2/3) \phi}(V(\varphi)+F)\right)=0 
                      \label{dilatonfull} ~,\\ \nn \\
\varphi: \;\;\; & & e^{-2 N_{0}}(\ddot{\varphi}-\dot{\varphi}
                         \dot{N_{0}}+3\dot{\varphi}\dot{R_{0}})+e^{(4/3) 
                         \phi_{0}} \frac{\dd V(\varphi)}{\dd \varphi}=0 
                         \label{scalarfull} ~,\\ \nn \\
\mbox{fluid}:\;\;\; & & p=p(\rho)~, \\  \nn \\
\mbox{00}: \;\;\; & & 3 e^{-2N}(\dot{R}^2 + \dot{R}\dot{B} - \frac{2}{9}
                    \dot{\phi}^2) + 3 e^{-2B}(-R''-2R'^2 + R'B'-\frac{2}{9}
                    \phi'^2) - \frac{1}{2}e^{(4/3) \phi}\La(\phi) \nn \\ 
                &-& \de(y)e^{-B}\left(\frac{1}{4}e^{-(2/3) \phi}e^{-2N}
                    \dot{\varphi}^2 
                    + \frac{1}{2}e^{(2/3) \phi}(V(\varphi)+\rho)
                    \right)=0 \label{00full}~, \\ \nn \\
\mbox{11}: \;\;\; & & e^{-2N}(-2\ddot{R}-\ddot{B}-3\dot{R}^2 - \dot{B}^2+2
                    \dot{N}\dot{R}+\dot{N}\dot{B}-2\dot{R}\dot{B}-\frac{2}{3}
                    \dot{\phi}^2) \nn \\  
                &+& e^{-2B}(N''+2R''+N'^2+3R'^2+2N'R'-N'B'-2R'B'+\frac{2}{3}
                    \phi'^2)+\frac{1}{2}e^{(4/3) \phi} \La(\phi)  \nn \\  
                &-& \de(y)e^{-B}\left(\frac{1}{4}e^{-(2/3) \phi}e^{-2N}
                    \dot{\varphi}^2 
                    + \frac{1}{2}e^{(2/3) \phi}(p-V(\varphi))\right)=0 ~,
                    \label{11full} \\ \nn \\
\mbox{05}: \;\;\; & & \dot{R}'+\dot{R}R' - N'\dot{R}-R'\dot{B}+\frac{4}{9}
                    \dot{\phi}\phi'=0~, \label{05full} \\ \nn \\
\mbox{55}: \;\;\; & & 3 e^{-2N}(-\ddot{R} - 2\dot{R}^2 + \dot{N}\dot{R} - 
                    \frac{2}{9}\dot{\phi}^2) + 3 e^{-2B}(N'R'+R'^2 -
                    \frac{2}{9}\phi'^2)+\frac{1}{2}e^{(4/3) \phi}\La(\phi) =0 
                    ~.	\label{55full}
\eea
In order to have a well defined geometry, the metric has to be 
continuous across $y=0$. However, first derivatives with respect 
to $y$ may not be continuous at $y=0$, and second derivatives may 
contain  delta functions. Such distributional parts can be treated 
separately by writing 
\be
\label{deltasplit}
f''=f''_{\mbox{\tiny{reg}}}+\de(y)[f']~,
\ee
where 
\be
[f'] \equiv \lim_{y\ra 0}\left(f'(y)-f'(-y)\right)
\ee
is the jump of $f'$ across $y=0$, and $f''_{\mbox{\tiny{reg}}}$ is the part which is 
regular at $y=0$. By matching the delta functions from the second
derivatives of  $\phi, N$ and $R$ with those in equations (\ref{dilatonfull}),
(\ref{00full}) and (\ref{11full}), one obtains the junction conditions
\be
[\phi'] = \frac{1}{8}e^{-(2/3)\phi_{0}} e^{B_{0} - 2 N_{0}} 
              \dot{\varphi}^2 + \frac{1}{4} e^{(2/3)\phi_{0}} 
              e^{B_{0}}(V(\varphi)+F) ~,
\ee
\be
[N']    = \frac{5}{12} e^{-(2/3)\phi_{0}} e^{B_{0} - 2 N_{0}} 
              \dot{\varphi}^2 + \frac{1}{6} e^{(2/3)\phi_{0}} 
              e^{B_{0}}(3p + 2 \rho - V(\varphi)) ~, \label{N'}
\ee
\be
[R']    = -\frac{1}{12} e^{-(2/3)\phi_{0}} e^{B_{0} - 2 N_{0}} 
              \dot{\varphi}^2 - \frac{1}{6} e^{(2/3)\phi_{0}} 
              e^{B_{0}}(V(\varphi)+\rho)~.  \label{R'}
\ee
Equations (\ref{N'}) and (\ref{R'}) are equivalent to  Israel's junction
conditions~\cite{Isra}. Our equations agree with those found by
other authors in special cases, see e.g. Refs.~\cite{BineI}
and~\cite{kss}. 

To ensure that the brane does not move, we  assume
$Z_2$ symmetry in the rest of this paper. Furthermore, we neglect
the dilaton and consider a simple cosmological constant in the bulk. For
the sake of generality these assumptions were not made so far.

\subsection{Special case: The Randall--Sundrum model}
The RS model is a special static solution of the equations derived in the 
previous section with $N(y)=R(y), B=0$, when $\La$ is taken to be a pure cosmological constant, and $V$ represents a constant brane tension. All other fields are set to zero. The RS metric is
\be
\label{backgroundmetric}
ds^2 = e^{2 \al |y|}(- dt^2 + d\vec{x}^2) + dy^2~.
\ee
Our equations of motion then reduce to
\bea
{\label{rs1} 
6 R'^2 = -\frac{\La}{2}} ~,\\
{\label{rs2}
3 R'' = -\de(y)\frac{V}{2}} ~.
\eea
Equation (\ref{rs1}) can now be solved by 
\be
\label{rssol}
R(y) = -\sqrt{\frac{-\La}{12}}|y| \equiv \al|y| ~,
\ee
which respects $Z_2$ symmetry and leads to an exponentially decreasing
`warp factor'.  To satisfy simultaneously equation (\ref{rs2}), one
must fine-tune the brane tension and the (negative) bulk cosmological
constant 
\be
\label{rsfine--tuning}
\La + \frac{V^2}{12}=0~.
\ee
This is the RS solution. A priori, $\La$ and $V$ are independent
constants, and there is no reason for such a relation. However, in a realistic
time-dependent cosmological model this relation must be satisfied in
order to recover the usual Friedmann equation for a fluid with
$\rho\ll V$ see Ref.~\cite{rsfried}. In the next section we study whether Eq.~(\ref{rsfine--tuning}) can be obtained by some dynamics on the brane.

\section{A dynamical brane}
\label{sec:dynamicalbrane}
We first consider a dynamical scalar field on the
brane.  The fine-tuning condition (\ref{rsfine--tuning})  corresponds to
the requirement that the negative bulk cosmological constant $\La$ can be
canceled by the brane tension $V$ which we try to identify with the
potential energy of  the  scalar field $\varphi$. If, starting with
some  initial conditions on $\varphi$ and $\dot{\varphi}$, the
evolution of the system would stabilize at $\La + \frac{V^2}{12}=0$,
the cancelation could be accomplished dynamically. If this would be the
case  for a `large class' of initial conditions, the RS solution
(\ref{rssol}) would be an attractor of the system. 

\begin{figure}[ht]
\centerline{\epsfxsize=3in  
\epsfbox{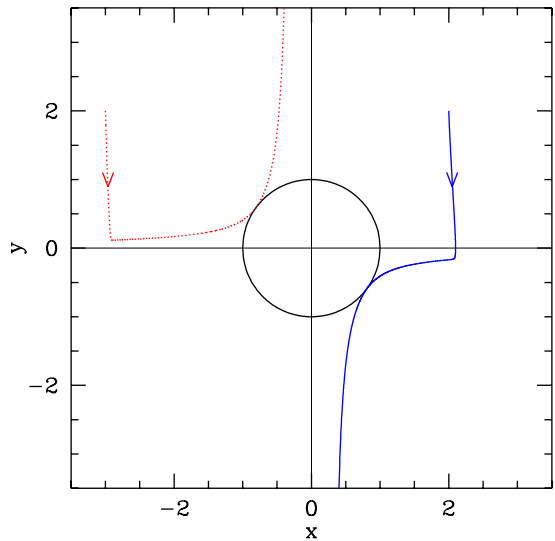}}
\caption{\label{rsca}Two trajectories in the phase space $(x, y)$
which represent typical solutions of the system
(\ref{system})-(\ref{constraint}) for $K=1$. The trajectory on the left
(dotted, red), starting with an initial condition 
$x_{\mbox{\tiny{in}}}=-3, y_{\mbox{\tiny{in}}}=2$,  
winds towards the circle $x^2+y^2 =\sqrt{K}$, which
corresponds to the condition $z=H=0$. After reaching the circle, the
solution moves away showing that it is not an attractor. The
trajectory on the right (solid) with initial conditions 
$x_{\mbox{\tiny{in}}}=2, y_{\mbox{\tiny{in}}}=2$
shows a similar behavior. It takes much longer to pass the region
around the kinks than to trace out the remaining parts of the trajectories.}
\end{figure}

\begin{figure}[ht]
\centerline{\epsfxsize=3in  
\epsfbox{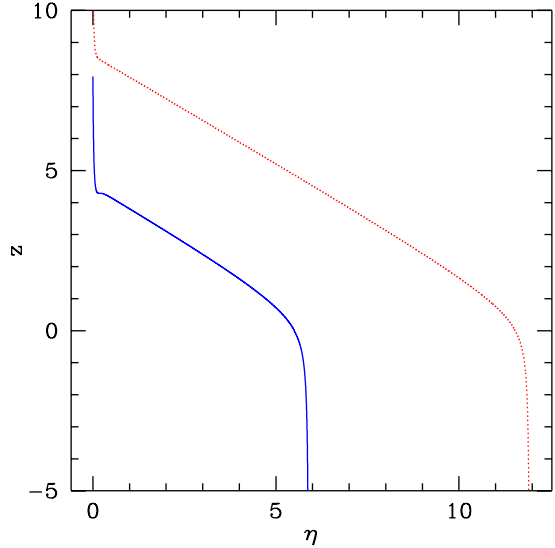}}
\caption{\label{figH}The time evolution of the dimensionless Hubble
parameter $z$ for the two trajectories shown in Fig.~1.}
\end{figure}

We start from Eqs. (\ref{scalarfull})-(\ref{55full}) for the case of a
vanishing dilaton. Taking the `mean value' of the $55$ equation 
across $y=0$, inserting the junction conditions (\ref{N'}), (\ref{R'})
and taking into account $Z_2$ symmetry, one obtains (see Ref.~\cite{BineI})
\be
\label{prefriedmann}
\ddot{R}_0 + 2 {\dot{R}_{0}}^2 = - \frac{1}{144} \rho_{b} (\rho_{b}
+3 p_{b}) + \frac{\La}{6},
\ee
where $\rho_{b}=\rho+\rho_{\varphi}$ and $p_{b}=p+p_{\varphi}$ are the total 
energy density and the total pressure on the brane due to the fluid  
and the scalar field. In this section the overdot denotes the derivative with
respect to the time coordinate $\tau$ given by $d\tau= e^{N_0(t)}dt$.
Using the energy conservation equation on the brane,
\be
\dot{\rho_{b}}=-3 \dot{R}_0 (\rho_{b}+p_{b}),
\ee
one can eliminate the pressure and integrate Eq. (\ref{prefriedmann})
to obtain a `Friedmann' equation for the expansion of the brane
(see Ref.~\cite{rsfried}) 
\be
\label{friedmann}
H^2 = \frac{1}{12} \La + \frac{1}{144} {\rho_{b}}^2 +
\frac{\cal C}{a_{0}^{4}},
\ee
where $a_{0}(t) \equiv e^{R(t,y=0)}$ denotes the scale factor on
the brane, $H=\dot{a}_0/a_0=\dot R_0$, and ${\cal C}$ is an integration constant. If the dilaton vanishes, 
Eq. (\ref{scalarfull}) becomes the ordinary equation of motion for a
scalar field 
\be
\label{scalar field}
\ddot{\varphi}+3H\dot{\varphi}+\frac{\partial V}{\partial \varphi}=0
\ee
with an energy density and  pressure
\bea
\rho_{\varphi}=\frac{1}{2}\dot{\varphi}^{2}+V(\varphi), 
\label{scalar fieldenergy} \\
p_{\varphi}=\frac{1}{2}\dot{\varphi}^{2}-V(\varphi).
\eea
We now assume that the energy density of the scalar field dominates any
other component on the brane, that is $\rho_{\varphi} \gg \rho$. This may be the case in the early universe. Later in this section we will see that this assumption does not affect our result. In the same sense we neglect the radiation term, so that Eq. (\ref{friedmann}) reduces to  
\be
\label{redfriedmann}
H = +\sqrt{\frac{1}{12} \left(\La + \frac{{\rho_{\varphi}}^2}{12} \right)}.
\ee
The positive sign corresponds to an expanding brane. The question of whether the system evolves towards $\La + \frac{V^2}{12}=0$ is now translated into the question of  whether the Hubble parameter vanishes at some time $\tau_1$.
From Eqs. (\ref{scalar field}) and (\ref{redfriedmann}) together with 
Eq. (\ref{scalar fieldenergy}) one finds
\be
\dot{H}=-\frac{1}{48} \rho_{\varphi} \dot{\varphi}^2,
\ee
which is always  negative. (The case $\dot{\varphi}(\tau_1)=0$
simultaneously with $H(\tau_1)=0$ will be treated separately.) Starting
with an expanding universe, $H>0$,   this implies that $H$ is indeed
decreasing and $H=0$ may well be obtained within  finite or infinite
time  depending on the details of the potential $V(\varphi)$. However,
at $\tau_1$ the scale factor has reached  a maximum  ($\ddot{a}_{0}(\tau_1)
=a_0(\tau_1)\dot H(\tau_1) <0$) and, after a momentary cancelation of $\La$
with  ${\rho_\varphi}^2$, $H$ changes sign and the brane
begins to contract 
with
\be
H = -\sqrt{\frac{1}{12} \left(\La + \frac{{\rho_{\varphi}}^2}{12}
\right)} ~.
\ee
In order for $H$ to stop
evolving at $\tau_1$ when the RS condition $\La+{\rho_\varphi}^2/12=0$ is
satisfied, we need $\frac{d^n}{d\tau^n}H(\tau_1)=0$ for all $n\ge 0$, which
implies $\frac{d^n}{d\tau^n}\rho_\varphi =0$ and also
$\frac{d^n}{d\tau^n} \varphi(\tau_1)=0$  for all $n\ge 1$.  
Therefore, the scalar field has to be  constant with value 
$\varphi_1 \equiv \varphi(\tau_1)$ and
$V(\varphi_1)=\sqrt{-12\La}$. But this is only possible if $V_1$
is a minimum of the potential, and we have to put
$\varphi$ into this minimum with zero initial velocity from the
start. This of course corresponds to the trivial static fine-tuned RS
solution. 
\\ \\
We conclude that the fine-tuning condition (\ref{rsfine--tuning}) cannot 
be  obtained  by such a mechanism.
Note that our arguments have been entirely general and we have thus
shown that the fine-tuning problem cannot be resolved by an arbitrary
brane scalar field.
\\ \\
To illustrate the dynamics, we consider the potential $V(\varphi)=\frac{1}{2}
m^{2} \varphi^{2}$.
Eq.~(\ref{redfriedmann}) then takes the form
\be
\label{examplefriedmann}
H^2 = \frac{1}{12} \La + \frac{1}{144}\left(\frac{1}{2} \dot{\varphi}^{2} 
+ \frac{1}{2} m^{2} \varphi^{2} \right)^{2}~.
\ee
It is convenient to use   dimensionless variables
$x$, $y$, $z$ and $\eta$ related to $\varphi$, $\dot{\varphi}$,
$H$ and $\tau$ by
\bea
\varphi \equiv \sqrt{\frac{24}{m}} x,  \;\;\; 
\dot{\varphi} \equiv \sqrt{24 m} y, \;\; \; 
H \equiv m z, \;\; \tau \equiv \frac{1}{m} \eta. 
\eea
Equations (\ref{scalar field}) and (\ref{examplefriedmann}) are equivalent 
to a two--dimensional dynamical system in the phase space $(x, y)$ with 
\be
\label{system}
x'=y, \;\;\;   y'=-x-3 z y, 
\ee 
with the constraint equation
\be
\label{constraint}
z^{2}=-K+(x^{2}+y^{2})^{2}.
\ee
The prime denotes the derivative with respect to the `time parameter'
$\eta$ and $K \equiv -\frac{1}{12 m^{2}} \La$. In Fig.~\ref{rsca} two typical
trajectories found by numerical solution of the system
(\ref{system})-(\ref{constraint}) are shown in the phase space $(x,y)$.
For large initial $y$, the damping term first dominates and lowers $y$
until the potential term becomes comparable. Then, the system evolves towards
the minimum of the potential until the curve hits the circle
$x^2+y^2=\sqrt{K}$ (after finite time), where the damping term changes sign, and the trajectories move away nearly in the $y$ direction. 
In Fig.~\ref{figH} the corresponding evolution of the Hubble parameter $z$ is shown. 
\vspace{1mm}
In ordinary four--dimensional cosmology there exists a `no--go theorem' due to Weinberg~\cite{Wei}, which states that the cosmological constant cannot be canceled by a scalar field. The argument is based on symmetries of the Lagrangian. 
In brane cosmology it is known~\cite{Shiro} that for a $3$--brane,  embedded in a five--dimensional spacetime, Einstein equations on the brane are the same as the usual four--dimensional Einstein equations, apart from two additional terms: A term  $S_{\mu\nu}$, 
which is quadratic in the energy--momentum tensor $T_{\mu\nu}$ 
on the brane  
 \begin{equation}
S_{\mu\nu} = -\frac{1}{4}T_{\mu\alpha}T_{\nu}^{~\alpha} 
+ \frac{1}{12}T T_{\mu\nu} 
+ \frac{1}{8} \ovl{g}_{\mu\nu} T_{\alpha \beta} T^{\alpha\beta}
- \frac{1}{24}\ovl{g}_{\mu\nu} T^2,
\end{equation}
and a term $E_{\mu\nu}$, which is the projection of the five--dimensional Weyl--tensor, ${C^A}_{BCD}$, onto the brane
\be
E_{\mu\nu} \equiv {C^A}_{BCD} n_A n^C \ovl{g}_{\mu}^{~B} \ovl{g}_{\nu}^{~D},
\ee
where $n_A$ is the normal vector to the brane and $T$ the trace of the energy 
momentum tensor. Being  constructed purely from $T_{\mu\nu}$,
$S_{\mu\nu}$ does not introduce additional dynamical degrees of
freedom. It just contributes the term ${\rho_\varphi}^2$ to the
`Friedmann' equation (\ref{friedmann}). It is also clear
that $E_{\mu\nu}$, which is traceless, cannot cancel the cosmological
constant on the brane. However, since
the effective Einstein equations on the brane cannot be derived from a
Lagrangian, and since $E_{\mu\nu}$ contains additional information from
the bulk, it is not evident that Weinberg's theorem holds in our
case.  
\\ \\
More generally, our `no-go' result also holds for any matter obeying an equation of state
$p=\omega \rho$ when $\omega > -1$. This can be seen in a similar way:
Initially the Hubble parameter is 
\be
H = +\sqrt{\frac{1}{12} \left(\La + \frac{\rho^2}{12} \right)}. \label{H}
\ee
Using the energy conservation equation
\be
\dot{\rho}=-3H(1+\omega)\rho,
\ee
one finds
\be
\dot{H}=-\frac{1}{48}(1+\omega)\rho^2, \label{Hdot}
\ee
and hence $\dot{H}<0$ as long as the weak energy condition, $\omega
>-1$ (or  $p>-\rho$) is satisfied.  To relate our finding to previous
results~\cite{rsfried,infla}, let us note that Eqs.~(\ref{H}) and (\ref{Hdot})
imply the following condition for inflation on the brane
\be {\ddot{a}_0\over a_0} = \dot H +H^2 = {\La\over 12} -{2+3\omega\over
144}\rho^2 > 0. \label{inf1}
\ee
For a brane energy density given by the brane RS tension $V=\sqrt{-12\La}$
and an additional component indicated by a subscript $_f$, so that
$\rho = V +\rho_f$ and $p=-V+p_f$ this gives
\be
 {\ddot{a}_0\over a_0} = -[V(1+3\omega_f)
		+\rho_f(2+3\omega_f)]{\rho_f\over 144} >0 ~,
\ee
which coincides with Eq.~(8) of Ref~\cite{infla}.
If the RS term dominates, $V\gg \rho_f$ we obtain the usual strong
energy condition
for inflation, $1+3\omega_f<0$, but if $V\ll \rho_f$ the condition is
stronger, namely, $2+3\omega_f<0$.

As in the case of  the scalar field, the
brane starts to contract as soon as $H=0$ is reached.
We have thus shown that a relation such as Eq. (\ref{rsfine--tuning}) cannot be
realized in a cosmological setting which does not violate the weak
energy condition. 

After this section, in which we adopted the viewpoint of the brane, we
now come back to the full five dimensional spacetime to investigate
the stability of the RS model.   
  
\section{Gauge Invariant Perturbation equations}
\label{sec:perturbationequations}
We formally prove that the five dimensional RS spacetime is unstable under small
perturbations of the brane tension.  
\subsection{Perturbations of the Randall--Sundrum model}
The equations of motion derived in Sec.~\ref{sec:equationsofmotion} provide with $N(t,y), R(t,y)$, and $B(t,y)$ a dynamical generalization of the RS model. We consider $\La$ and $V$ to be constant and set the dilaton, the scalar field on the brane and the energy-momentum tensor of the fluid to zero.
Equations (\ref{00full})-(\ref{55full}) now reduce to
\bea
\mbox{00}: \;\;\; & & 3 e^{-2N}(\dot{R}^2 + \dot{R}\dot{B}) + 3 e^{-2B}(-R''-
                    2R'^2 + R'B') - \frac{1}{2}\La - \de(y) 
                    \frac{1}{2} e^{-B} V = 0~, 
                    \label{00rstime} \\
\mbox{11}: \;\;\; & & e^{-2N}(-2\ddot{R}-\ddot{B}-3\dot{R}^2 - \dot{B}^2+
                    2\dot{N}\dot{R}+\dot{N}\dot{B}-2\dot{R}\dot{B}) \nn \\
                &+& e^{-2B}(N''+2R''+N'^2+3R'^2+2N'R'-N'B'-2R'B')+
                    \frac{1}{2}\La + \de(y) 
                    \frac{1}{2} e^{-B} V = 0~, \label{11rstime}\\
\mbox{05}: \;\;\; & & \dot{R}'+\dot{R}R' - N'\dot{R}-R'\dot{B}=0~, 
                    \label{05rstime} \\
\mbox{55}: \;\;\; & & 3 e^{-2N}(-\ddot{R} - 2\dot{R}^2 + \dot{N}\dot{R}) + 
                    3 e^{-2B}(N'R'+R'^2)+\frac{1}{2}\La =0~. \label{55rstime}
\eea
The RS solution (\ref{rssol}) is a static solution of these equations, provided that condition (\ref{rsfine--tuning}) holds. We now derive linear perturbation equations from Eqs.(\ref{00rstime})--(\ref{55rstime}) which describe  the
time evolution of small deviations from RS. To this goal we set
\bea
N(t,y) &=& \al |y| + n(t,y), \\
R(t,y) &=& \al |y| + r(t,y), \\
B(t,y) &=& b(t,y),
\eea
where $\al = -\sqrt{\frac{-\La}{12}}$ and $n(t,y), r(t,y), b(t,y)$ are
small at $t=0$. The perturbed metric is
\be
\label{perturbedmetric}
ds^2 = -e^{2 \al |y| + 2 n} dt^2 
+ e^{2 \al |y| + 2 r} d\vec{x}^2 + e^{2 b} dy^2.
\ee
We consider an energy-momentum tensor deviating from RS only by a
slight mismatch of the brane tension
\be
\label{energymomentum}
T_{MN}=-\La g_{MN} - \de(y) \de^{\mu}_{M} \de^{\nu}_{N} e^{-b} V 
       \ovl{g}_{\mu \nu},
\ee 
with 
\be
V=\sqrt{-12 \La}(1+\Om),
\ee
where $|\Om|\ll 1$ parameterizes the perturbation of the brane tension,
$g_{MN}$ is the perturbed metric~(\ref{perturbedmetric}) and 
$\ovl{g}_{\mu \nu}$ is its projection onto the brane. 
Clearly, if already this restricted set of perturbation variables
contains an instability, the RS solution is unstable
under homogeneous and isotropic perturbations.
Inserting this ansatz into equations (\ref{00rstime})--
(\ref{55rstime}) and keeping only first order terms, we find 
\bea
& & r'' - 4 \al^2 b + \th(y) \al(4 r'-b')-\de(y)2 \al(b+\Om)=0 ~,
    \label{00per} \\
& & e^{-2 \al |y|}(2 \ddot r + \ddot b)-n''-2r''+12 \al^2 b 
    - \th(y) \al(4 n'+8 r'-3 b')+ \de(y)6 \al (b+\Om)=0 ~,
    \label{11per} \\
& & \dot r' - \th(y) \al \dot b =0 ~,\label{05per} \\
& & e^{-2 \al |y|} \ddot r + 4 \al^2 b - \th(y) \al(n'+ 3  r')=0~, 
    \label{55per}
\eea
where
\be
\th(y) = \left\{\begin{array}{ll}
      +1 & \mbox{for $y>0$} \\
      -1 & \mbox{for $y<0$}~.
      \end{array}
\right. 
\ee
The  junction conditions are
\be
[r']=[n']=2 \al(b_0+\Om)~.
\ee
Since we want to consider $Z_2$-symmetric perturbations, we require
the functions $n$, $r$ and $b$ to be symmetric in $y$.
In order to  make coordinate-independent statements, we rewrite these 
equations in a gauge invariant way. 
\subsection{Gauge invariant perturbation equations}
Under an infinitesimal coordinate transformation induced by the vector field
\be
X=T(t,y)\partial_t + L(t,y)\partial_y,
\ee
the metric perturbations $g^{(1)}$ transform according to 
\be
g^{(1)} \rightarrow g^{(1)} + {\mathcal{L}}_X g^{(0)},
\ee
where $g^{(1)}$ corresponds to the first order terms in the 
metric (\ref{perturbedmetric}), and ${\mathcal{L}}_X g^{(0)}$ is the
Lie derivative of the static background metric (\ref{backgroundmetric}).
One obtains the following transformation laws for the variables $n$,
$r$ and $b$: 
\bea
& & n \rightarrow n + \th(y) \al L + \dot{T},  \label{n.gaugetf} \\
& & r \rightarrow r + \th(y) \al L, \label{a.gaugetf} \\
& & b \rightarrow b + L'. \label{b.gaugetf}
\eea
Since we require the 05 component of the metric to vanish, it must remain zero under the coordinate transformation. This implies
\be
\label{LTlink}
\dot{L}=e^{2 \al |y|} T'.
\ee
From Eq.~(\ref{b.gaugetf}), together with $Z_2$ symmetry, one finds
that $L'$ must be continuous and symmetric in $y$. Therefore $L$ must
be continuously differentiable and odd in $y$, which implies $L(t,y=0)=0$.  
Hence, the perturbation $r$ restricted to the brane $r_0$ is gauge
invariant. Note that $L(t,y=0)=0$ also follows from
Eq.~(\ref{a.gaugetf}) and $L\in C^1$. Hence the gauge invariance of
$r_0$ is not a consequence of $Z_2$ symmetry, but is also preserved
for non--$Z_2$ symmetric perturbations.
By computing the Lie derivative of the background energy-momentum
tensor from Eq.~(\ref{energymomentum}) one finds that the perturbation
of the brane tension $\Om$  is  gauge invariant.  Condition 
~(\ref{LTlink}) and the symmetry property of $L'$ ensure that there is no 
energy flow  onto or off the brane. With the following set of gauge
invariant quantities  
\bea
& & \Phi \equiv r'-\th(y) \al b ~,\\
& & \Psi \equiv n'-\th(y) \al b - \th(y) \frac{1}{\al} 
    e^{-2 \al |y|} \ddot{r}~, \\
& & r_0 \equiv r(t,y=0)~, \\
& & \Om,
\eea
we can rewrite the perturbation equations (\ref{00per})--(\ref{55per}) in
terms of these variables 
\bea
& & \Phi'+ \th(y)4 \al \Phi - \de(y) 2 \al \Om = 0 ~,
    \label{00per.eiv}\\
& & \Psi' + 2 \Phi' + \th(y) 4 \al (\Psi+2\Phi) + 
    \de(y) \left(\frac{2}{\al}\ddot{r}_0 -6 \al \Om \right) = 0~,
    \label{11per.eiv}\\
& & \dot{\Phi}=0 ~,
    \label{05per.eiv} \\
& & \Psi+3\Phi=0 ~.
    \label{55per.eiv} 
\eea
The junction conditions are
\be
[\Phi]=2 \al \Om ~,\qquad \qquad 
[\Psi]=2 \al \Om - \frac{2}{\al} \ddot{r}_0~.
\ee
The solutions of equations (\ref{00per.eiv}) with (\ref{05per.eiv})
and  (\ref{55per.eiv}) are given by
\be
\Phi(y)= -\frac{1}{3}\Psi(y) = \th(y) \al \Om e^{-4 \al |y|}
         + \Phi_0 e^{-4 \al |y|}~.
\ee
$Z_2$ symmetry requires $\Phi$ to be odd in  $y$ and thus $\Phi_0=0$. 
Inserting Eq. (\ref{55per.eiv}) in (\ref{11per.eiv})
one obtains
\be
\ddot{r}_0 = 4 \al^2 \Om
\ee
and after integration 
\be
r_0(t)=2 \al^2 \Om t^2 + {\cal Q} t,
\ee
where ${\cal Q}$ is a small but arbitrary integration constant
determined by the initial conditions. (An additive constant to
$r_0$ can be absorbed in a redefinition of the spatial coordinates on the 
brane.)
The scale factor on the brane is 
\be
e^{2 r_0(t)} \simeq 1 + 2 r_0(t)=1+4 \al^2 \Om t^2 + 2 {\cal Q} t.
\ee
We have thus found a dynamical instability, which is quadratic in time,
when the brane tension and the bulk cosmological constant are not
fine--tuned. Our  statement is valid in every coordinate 
system as  $r_0$ is gauge invariant. In addition,  more surprisingly, 
in linear perturbation theory there is no constraint on ${\cal Q}$, 
and it cannot be gauged away. This linear instability remains even 
for $\Om = 0$, that is  if the  brane tension  is not perturbed at all. 
\\ \\
Let us finally discuss our solutions in two particular gauges.
As a  first gauge condition we set $r'=0$, which  fixes 
$L'=-\th(y)\frac{1}{\al} r'$. The integration constant on $L$ is 
determined by the condition $L(t,y=0)=0$. (Note that $r'$ contains 
a $\th$ function and therefore $L'$ is continuous.) For all values 
of $y$ we have
\be
r(t)=2 \al^2 \Om t^2 + {\cal Q} t.
\ee
Since $b(y)=-\th(y)\frac{1}{\al} \Phi(y)$,
\be
b(y)=-\Om e^{-4 \al |y|}.
\ee
From the definition of $\Psi$ it follows that
\be
n'= - 3 \Phi + \th(y) \al b + \th(y) \frac{1}{\al} e^{-2 \al |y|} \ddot{r}_0~, 
\ee
which can be integrated to give
\be
n(t,y)=\Om e^{-4 \al |y|} - 2 \Om e^{-2 \al |y|} +{\cal N}(t)~.
\ee
These $n, r$, and $b$ solve Eqs. (\ref{00per})--(\ref{55per}).
The integration constant ${\cal N}(t)$ can be absorbed in the gauge
transformation  $T$. Together with the choice of $L'$, this fixes the gauge, and the solutions are therefore unique up to an additive purely time dependent function to $T$.  
\\ \\
Another possible gauge is $b=0$. Then, from  $r'=\Phi$,
\be
r(t,y)=-\frac{\Om}{4}e^{-4 \al |y|} + {\cal R}(t),
\ee
with 
\be
{\cal R}(t)=r_0(t)+\frac{\Om}{4}=2 \al^2 \Om t^2 + {\cal Q} t
+\frac{\Om}{4} 
\ee
and 
\be
n(t,y)=\frac{3}{4} \Om e^{-4 \al |y|} - 2 \Om e^{-2 \al |y|}
 +{\cal N}(t)~.
\ee
Again, the integration constant ${\cal N}(t)$ can be gauged away by 
choosing an appropriate $T$, and the solutions are uniquely 
determined by the gauge fixing. 
\\ \\
Inserting these solutions in the perturbed metric 
\be
\label{pertubedmetric}
ds^2 = -e^{2 \al |y| + 2 n} dt^2 
+ e^{2 \al |y| + 2 r} d\vec{x}^2 + e^{2 b} dy^2,
\ee
we find that the full RS spacetime, not only the brane, is unstable
against homogeneous perturbations of the brane tension.  

We must require the initial perturbations to be small, that is at some
 initial time $t=0$, the deviation from $RS$ has to be small for all
 values of $y$. In the case of a compact spacetime $|y|\le y_{\max}$,
 this just requires  $|\Om| \ll  e^{-4\al y_{\max}}$ (remember that
 $\al$ is a negative constant). For 
a noncompact  spacetime $-\infty <y<\infty$, we have to require
$\Om=0$. In other words, for $V\not\equiv \sqrt{-12\La}$ there
exists no solution which is `close' to RS in the sense of $L^2$
or $\sup_y$ at any given initial time.
\\ \\
Finally, we present a geometrical interpretation of the 
gauge invariant quantities $\Phi$ and $\Psi$.   Since the five--dimensional
Weyl tensor of the RS solution vanishes, the perturbed Weyl tensor is
gauge invariant according to the Steward-Walker lemma~\cite{StWa}.
The 0505 component of the Weyl tensor of the perturbed metric
(\ref{perturbedmetric}) is up to first order  
\be
C_{0505}=\frac{1}{2} e^{2 \al |y|} (n''-r'' +\th(y) \al (n'-r')) 
+\frac{1}{2}(\ddot{r}-\ddot{b}),
\ee
which can be expressed in terms of gauge invariant quantities
\be
C_{0505}=-\frac{1}{2} e^{\al |y|}\left(e^{\al |y|}(\Phi-\Psi)\right)' 
	 + \de(y) \frac{1}{\al} \ddot{r}_0.
\ee
All other nonvanishing Weyl components are multiples of $C_{0505}$:
\bea
& & C_{0101}=C_{0202}=C_{0303}=C_{1212}=C_{1313}=C_{2323}=
-\frac{1}{3} e^{2 \al |y|} C_{0505}  ~~\mbox{ and}\\
& & C_{1515}=C_{2525}=C_{3535}=\frac{1}{3} C_{0505}~.
\eea
In first order the projected Weyl tensor (defined in Ref.~\cite{Shiro}) is 
$E_{11}=E_{22}=E_{33}=\frac{1}{3}E_{00}=2\al^2\Om$ with ${E^{\mu}}_{\mu}=0$.
The Weyl--tensor completely vanishes for $\Om=0$. 
\section{Results and Conclusions}
In this paper we have addressed two main questions: First, we investigated
whether the RS fine--tuning condition can be obtained dynamically by some 
matter
component on the brane. As a concrete example, we studied a scalar
field on the brane  and found that a bulk
cosmological constant cannot be canceled by the potential of the
scalar field in a nontrivial way. This result can be generalized for
any matter satisfying the weak energy condition. 

Second, we studied the stability of the RS model in five dimensions.
We have found that the RS solution  is unstable under
homogeneous and isotropic, but time dependent perturbations. For a
small deviation of the 
fine--tuning condition parametrized by $\Om\neq 0$, this 
instability was expected. It is very reminiscent of
the instability of the static Einstein universe, where the fluid
energy density and the cosmological constant have to satisfy a
delicate balance in order to keep the universe static. But even if
$\Om=0$, there exists a  mode (${\cal Q} \neq 0$), which 
represents an instability in first order perturbation
theory. It is interesting to note that in a non--$Z_2$ symmetric setting this 
mode can be absorbed into a motion of the brane. 
In a cosmological context, our result means that a possible change in
the brane tension,  e.g. during a phase transition, or also quantum
corrections to the bulk energy density  (see Ref.
\cite{quantumcorrections}) could give rise to instabilities of
the full five--dimensional  spacetime. For related work see Refs.~\cite{kim},
\cite{kaloper}, \cite{shanki}.

Even if one would consider a dynamical scalar field in the bulk (which
does not couple to brane fields), which settles into a vacuum state
such that its energy density is constant along the fifth dimension,  
one would not be able to solve the cosmological constant problem 
without falling back to some fine--tuning mechanism. From our results
we can conclude that in order to have a chance
to solve the RS fine tuning problem dynamically, we have to consider
fully dynamical {\em bulk} fields. This can in principle be done with
the system of equations, which we have presented in Sec.
\ref{sec:equationsofmotion}, and which also applies to the effective
five--dimensional low--energy theory  suggested by heterotic
M theory~\cite{Lukas}.  

\vspace{10mm}
{\bf Acknowledgments} 
\\ \\ We wish to thank Arthur Hebecker, Kerstin Kunze, Marius Mantoiu 
Dani\`ele Steer, Toby Wiseman and Peter Wittwer for useful discussions and comments. C.v.d.B thanks Geneva
university for hospitality. This work was
supported by the Swiss National Science Foundation.
C.v.d.B was supported by the Deutsche Forschungsgemeinschaft (DFG).


\end{document}